\documentstyle[11pt,aaspp4]{article}
\begin{document}

\def\Ej{{\rm E}_{\rm J}}
\def\ring{{\rm ring}}
\def\bar{{\rm bar}}
\def\ering{{\rm eff,ring}}
\def\ebar{{\rm eff,bar}}

\title{DYNAMICAL EFFECTS OF NUCLEAR RINGS \\
       IN DISK GALAXIES}

\author{Clayton H. Heller}
\affil{Universit\"ats Sternwarte \\
       Geismarlandstra\ss e 11 \\
       D-37083 G\"ottingen, Germany \\
       Email: cheller@uni-sw.gwdg.de}
\and
\author{Isaac Shlosman}
\affil{Department of Physics and Astronomy \\
       University of Kentucky \\
       Lexington, KY 40506-0055, USA \\
       Email: shlosman@asta.pa.uky.edu}

\begin{abstract}

We investigate the dynamical response of stellar orbits in a rotating
barred galaxy potential to the perturbation by a nuclear gaseous ring. The 
change in 3D periodic orbit families is examined as the gas accumulates
between the inner Lindblad resonances. It is found that the phase space 
allowable to the $x_2$ family of orbits is substantially increased and a 
vertical instability strip appears with the growing mass of the ring. A 
significant distortion of the $x_1$ orbits is observed in the vicinity of 
the ring, which leads to the intersection between orbits with different 
values of the Jacobi integral. We also examine the dependence of the orbital 
response to the eccentricity and alignment of the ring with the bar.
Misalignment between an oval ring and a bar can leave observational 
footprints in the form of twisted near-infrared isophotes in the vicinity
of the ring.  It is inferred that a massive nuclear ring acts to weaken
and dissolve the stellar bar exterior to the ring, whereas only weakly
affecting the orbits interior to the inner Lindblad resonances. 
Consequences for gas evolution in the circumnuclear regions of barred 
galaxies are discussed as well.
                          
\end{abstract}


\newpage

\section{INTRODUCTION}

It is generally agreed that barred disk galaxies experience radial inflows 
of the interstellar medium (ISM) (e.g. Kenney \markcite{ken96} 1996,
and refs.  therein). These flows are presumably triggered by gravitational
torques from non-axisymmetric background potentials, such as stellar bars
and ovals. (For a more comprehensive discussion of this phenomenon see
review by Shlosman, Begelman \& Frank \markcite{shl90} 1990).  A large 
fraction, $\sim2/3$, of disk galaxies are observed to be barred to 
various degrees (e.g. de Vaucouleurs \markcite{dev63} 1963), the remaining 
1/3 typically consist of edge-on galaxies where bars would be difficult to 
detect (e.g. review by Martinet \markcite{mar95} 1995). In addition, most 
galactic disks are ovally distorted (Bosma \markcite{bos81} 1981; Kormendy 
\markcite{kor82} 1982; Rix \& Zaritsky \markcite{rix95} 1995), especially
in the central regions where triaxial stellar bulges are frequent 
(Kormendy \markcite{kor93} 1993). 
 
Theoretical studies of gaseous inflows confirm that gas can reach the 
circumnuclear regions of disk galaxies (e.g. Combes and Gerin \markcite{com85}
1985; Shlosman, Frank \& Begelman \markcite{shl89} 1989; Athanassoula 
\markcite{ath92} 1992; Friedli and Benz \markcite{fri93} 1993;
Shlosman and Noguchi \markcite{shl93} 1993) and cause enhanced star 
formation there, so-called nuclear starbursts (Heller and Shlosman 
\markcite{hel94} 1994; Knapen etal. \markcite{kna95a} 1995a). 
Indeed, observations of nuclear starburst galaxies 
reveal ring-like molecular gas accumulations within the central kpc (e.g. 
Telesco and Decher \markcite{tel88} 1988; Pogge \markcite{pog89} 1989; 
Kenney etal. \markcite{ken92} 1992; Benedict etal. \markcite{ben93} 1993;
Knapen etal. \markcite{kna95b} 1995b), where
a slowdown of radial inflow is expected in the presence of inner Lindblad
resonance(s) (ILRs) (Combes \& Gerin \markcite{com85} 1985; Shlosman, 
Frank \& Begelman \markcite{shl89} 1989). 
Nuclear rings are ovally shaped and their major axes are
observed to lead the stellar bars, typically by $\sim 50\arcdeg-90\arcdeg$
in the direction of galactic rotation. Frequently, they are patchy and
incomplete (Buta \& Crocker \markcite{but93} 1993). A substantial fraction, 
$\sim10\%-30\%$, of the molecular galactic ISM can accumulate in these rings, 
which consist of a mixture of neutral and ionized, gas and dust, and is
accompanied by massive star formation. Ring masses can be roughly estimated
using the axisymmetric rotation curve and/or emission from this gas. 
Typical masses range from few$\times 10^8\,M_{\sun}$ to few$\times 
10^9\,M_{\sun}$, and even higher (Kenney \markcite{ken96} 1996 and 
refs. therein). Besides being an active site of star formation, 
a nuclear ring provides a significant gravitational perturbation to the
background galactic potential, and is expected to alter the main stellar
and gas orbits in the central regions. Growing molecular rings further
contribute to the increased mass concentration in the disk. Because of 
this we expect them to modify the inner galactic rotation curve and the
positions and strength of the ILRs.

Previous studies of stellar orbits in a rapidly rotating barred potential
focused on the secular effects of a growing {\it spherically-symmetric} 
central mass concentration, e.g. massive black hole and/or galactic bulge,
exceeding $10^9\,M_{\sun}$, or $\sim1\%$ of the overall mass of a galaxy
(Martinet \& Pfenniger \markcite{mar87} 1987; Hasan \& Norman \markcite{has90} 
1990; Hasan, Pfenniger \& Norman \markcite{has93} 1993). In particular, the 
main 3D periodic orbits in a barred system with a central
mass have been identified and their stability analyzed. These studies exposed
the crucial role of ILRs and higher-order resonances in increasing the 
stochasticity in the system, or simply heating it up, which results in 
the weakening and dissolution of the stellar bar and the buildup of the
galactic bulge.
This interesting phenomenon has led the authors to argue in favor of galaxy 
evolution along the Hubble sequence (Pfenniger, Combes and Martinet
\markcite{pfe94} 1994; Martinet \markcite{mar95} 1995). 
Implications for gas dynamics have been studied by Pfenniger \& Norman 
\markcite{pfe90} (1990) neglecting the self-gravity in the gas.

Although it is possible that all nuclear starburst galaxies (and even all disk 
galaxies!) harbor central black holes (BHs), it is by no means clear
that the masses of these BHs are in the range of $\gtrsim10^9\,M_{\sun}$. 
It is plausible that typical BHs in disk galaxies, active and normal, are much 
less massive, and lie in the mass range of $\sim10^7-10^8\,M_{\sun}$
(with the exception of quasar hosts). For example, Emmering, Blandford and
Shlosman \markcite{emm92} (1992) have argued that BHs in Seyfert nuclei 
may be more efficient in extracting energy and angular momentum from the
accreting material than commonly believed, allowing for lower estimates of BH
masses there.  On the other hand, a buildup of galactic bulges may proceed on 
a timescale which is a fraction of a Hubble time and cause the same effects 
on the galactic dynamics within the bar region, as discussed in the previous 
paragraph.

Evidently, the estimated masses of nuclear rings make them competitive with 
galactic bulges in affecting the structure of disk galaxies. At the same time,
their morphology differs quite dramatically from that of nearly
spherically-symmetric or weakly triaxial bulges. This motivated us to 
analyze the main periodic orbits in a barred potential perturbed by a 
growing nuclear ring. We aim at understanding the gravitational effects
of this phenomenon on the stellar orbits, as well as deduce possible 
implications on the gas evolution in these galaxies. 
Both circular and oval rings are used, and the results
are compared to the previous studies of such orbits, with and without
the central mass (e.g. Contopoulos \& Papayannopoulos \markcite{con80} 1980; 
Athanassoula etal. \markcite{ath83} 1983; Pfenniger \markcite{pfe84} 1984;
Hasan, Pfenniger \& Norman \markcite{has93} 1993). We find that in many 
ways, the dynamical consequences of nuclear rings differ from those of
a spherically-symmetric central mass, as addressed so far in the literature.

In the next section we describe a model for the background galactic
potential, which is perturbed by a massive ring. The potential of an elliptical
hoop, representing the ring, is calculated analytically. The third section
presents the method we employ to locate the main families of periodic 
orbits in the phase space. Readers that wish to avoid the technical details
may go directly to the fourth section, where the results of the orbital
analysis are given for a series of circular and elliptical ring models,
with gradually increasing mass and different orientations with respect 
to the stellar bar. A discussion of the possible effects on the gas flow
in a galaxy is given in the last section. 

\section{MODEL}

The galaxy model consists of the superposition of four components:
disk, bulge, bar, and ring.  For the disk we choose a potential introduced 
by Miyamoto and Nagai \markcite{miy75} (1975),
\begin{equation}
\Phi_{\rm d} = -\frac{G M_{\rm d}}{\sqrt{R^2 + \left(A_{\rm d} + 
		\sqrt{B_{\rm d}^2 + z^2} \, \right)^2 }},
\end{equation}
where $R^2 = x^2 + y^2$.  This axisymmetric potential is characterized by the
total mass $M_{\rm d}$, horizontal scale $A_{\rm d} + B_{\rm d}$,
and vertical scale $B_{\rm d}$.  When $B_{\rm d}= 0$, $\Phi_{\rm d}$
reduces to Kuzmin's very flattened potential.
For the spherically symmetric bulge or spheroidal component we will use 
a Plummer sphere,
\begin{equation}
\Phi_{\rm s} = -\frac{G M_{\rm s}}{\sqrt{R^2 + z^2 + B_{\rm s}^2}}.
\end{equation}

The bar is represented by a triaxial Ferrers \markcite{fer77} (1877) 
density distribution of semi-axes $a$, $b$, and $c$, given by
\begin{mathletters}
\begin{equation}
\rho = \left\{ \begin{array}{ll}
             \frac{105 M_{\rm b}}{32 \pi a b c} \left(1 - m^2 \right)^2 
             & \quad \mbox{if} \quad m<1 \\ 0 & \quad \mbox{if} \quad m \ge 1,
	       \end{array} \right.
\end{equation}
where
\begin{equation}
m^2 = \frac{x^2}{a^2} + \frac{y^2}{b^2} + \frac{z^2}{c^2}
\end{equation}
\end{mathletters}
and $a>b>c$. We have adopted the strict inequality, rather than use
a prolate distribution, because of the desire to represent an intermediate 
strength bar of moderate eccentricity, as is observed in many of the
galaxies which are known to contain nuclear rings of gas.  We use
the technique of Pfenniger \markcite{pfe84} (1984) to evaluate the 
corresponding potential.

The ring is represented by the fully 3D softened potential of a uniform density
elliptical curve in the disk plane of linear density $\lambda$, as given by,
\begin{equation}
\Phi_{\rm r} = -G\lambda\int\!\frac{ds}{d},
\end{equation}
where $d$ is the distance between a point on the ring $[r,\phi,0]$ and
the point of evaluation $[R,\theta,z]$. With the functions
\begin{mathletters}
\begin{equation}
g(\phi) = \sqrt{\frac{1-e^2}{1-e^2\cos\phi^2}}
\end{equation}
and
\begin{equation}
f(\phi) = \frac{e^2\sin\phi\cos\phi}{1-e^2\cos^2\phi},
\end{equation}
\end{mathletters}
the potential can be written as
\begin{equation}
\Phi_{\rm r} = -G\lambda R_0 \int_\theta^{\theta + 2\pi}
	       \frac{g(\phi) \sqrt{1 + f(\phi)^2} \ d\phi}
	       {\sqrt{R^2 + z^2 + R_0^2 g(\phi)^2 - 2 R R_0 g(\phi) 
	       \cos(\phi-\theta) + \epsilon^2}},
\end{equation}
where $R_0$ is the semi-major axis, $e$ the ellipticity, and $\epsilon$
a softening parameter which gives the ring an effective thickness.
Expanding the integrand in $e$ to second order and substituting
$\phi = \pi - 2\psi + \theta$ gives,
\begin{equation}
\Phi_{\rm r} = -G\lambda R_0 \int_0^{\pi/2} \left[ \frac{C_0}{\Delta} + e^2 
       \frac{C_1 + C_2\cos^2\psi + C_3\cos^4\psi + C_4\cos^6\psi}{\Delta^3}
       + O(e^4) \right] d\psi,
\end{equation}
where,
\begin{mathletters}
\begin{eqnarray}
\Delta   & = & \sqrt{1-k^2\sin^2\psi}, \\
k        & = & \sqrt{1 - k'^2}, \\
k'       & = & R_1/R_2, \\
R_1^2 & = & (R-R_0)^2 + z^2 + \epsilon^2, \\
R_2^2 & = & (R+R_0)^2 + z^2 + \epsilon^2,
\end{eqnarray}
\end{mathletters}
and the coefficients, which depend only on the point of evaluation,
are given by
\begin{mathletters}
\begin{eqnarray}
C_0 & = & -4, \\
C_1 & = & (2\,R_0{x}^{2}R-2\,{R}^{3}R_0-2\,{\epsilon}^{2}{x}^{2}+
	  2\,{R}^{4}+2\,{z}^{2}{R}^{ 2}-2\,{x}^{2}{R}^{2} \nonumber \\
    &   & -2\,{z}^{2}{x}^{2}+2\,{\epsilon}^{2}{R}^{2})/{R}^{2}, \\
C_2 & = & -(8\,{R}^{4}+8\,{\epsilon}^{2}{R}^{2}-16\,{\epsilon}^{2}
       {x}^{2}+8\,{z}^{2}{R}^{2}-12\,{R}^{3}R_0-16\,{z}^{2}{x}^{2} \nonumber \\
    &   & -16\,{x}^{2}{R}^{2}+20\,R_0{x}^{2}R)/{R}^{2}, \\
C_3 & = & (-16\,{z}^{2}{x}^{2}-16\,{x}^{2}{R}^{2}+8\,{R}^{4}-16\,
	  {\epsilon}^{2}{x}^{2}+8\,{z}^{2}{R}^{2}-24\,{R}^{3}R_0 \nonumber \\
    &   & +48\,R_0{x}^{2}R+8\,{\epsilon}^{2}{R}^{2})/{R}^{2}, \\
C_4 & = & -\frac{16\,R_0\left (2\,{x}^{2}-R^2\right )}{{R}}. 
\end{eqnarray}
\end{mathletters}

This may now be integrated using standard integral tables to give
\begin{equation}
\Phi_{\rm r} = \frac{G\lambda R_0}{R_2} \left[ C_0 I_0 + 
	\frac{e^2}{R_2^2} \left( C_1 I_1 + C_2 I_2 + C_3 I_3 + 
	C_4 I_4 \right) + O(e^4) \right],
\end{equation}
with,
\begin{mathletters}
\begin{eqnarray}
I_0 & = & K(k), \\
I_1 & = & E(k)/k'^2, \\
I_2 & = & \frac{K(k) - E(k)}{k^2}, \\
I_3 & = & \frac{(k'^2+1) E(k) - 2 k'^2 K(k)}{k^4}, \\
I_4 & = & \frac{(2 k^2-4 k'^2) I_3 + 3 k'^2 I_2}{3 k^2},
\end{eqnarray}
\end{mathletters}
where $K$ and $E$ are the complete elliptic integrals of the 1st and 2nd
kinds, respectively.

For $e=0$ the above expression reduces to the the 3D potential of a circular
hoop.  Substituting $M_{\rm r} = 2\pi\lambda R_0$ this gives
\begin{equation}
\Phi_{\rm hoop} = -\frac{2 G M_{\rm r}}{\pi R_2} K(k)
\end{equation}
and can be compared with the derivation by MacMillan \markcite{mac58} (1958).
The full expression has been checked against an numerical integration 
of Equation~6.

For the model parameters we have chosen values of $M_{\rm d} = 1.7$,
$A_{\rm d} = 0.5$, $B_{\rm d} = 0.05$, $M_{\rm s} = 0.1$, 
$B_{\rm s} = 0.05$, $M_{\rm b} = 0.12$, $a = 0.5$, $b = 0.25$, 
$c = 0.05$, $R_0 = 0.04$, and $\epsilon = 0.01$. With adopted
units for mass and distance of $M = 10^{11}\,M_{\sun}$ and $R=10$\,Kpc,
respectively, this model has a mass within 10\,Kpc of approximately
$1.1\!\times\!10^{11}\,M_{\sun}$.  The bar of length 5\,Kpc and
axial ratios $b/a = 0.5$ and $c/a = 0.1$, represents about 22\% of
the mass within its radius.  The center of the ring is at a radius (or a 
semi-major axis) of 
$R_0=400$\,pc in the disk plane and is smeared by the softening so that about 
60\% of its mass is contained in a cylindrical shell of 200\,pc 
in radius and height centered at $R_0$. This was done to avoid singular
forces at the position of the ring.
We will examine models with ring masses of 0.0, $10^8$, 
$5\!\times\!10^8$, and $10^9$ solar masses, which are added to the mass
of the system. Because the mass of the ring is insignificant compared to the
overall mass, and because the system itself is not dynamically self-consistent, 
this is the simplest way to account for the gas accumulation in the ring.
These rings represent, respectively, 0, 6, 23, and 38 percent of the mass
inside the cylindrical shell (for circular rings).  The bar pattern speed was 
chosen for the corotation to be at the end of the bar, which gives 
$\Omega_{\rm p} = 2.284$.  Two radial (and one vertical) ILRs exist at 
this value of $\Omega_{\rm p}$, as discussed in Section 4.
The first four models (A--D)  have either none or a circular ring ($e=0$),
the next three (E--G) have an elliptical ring ($e=0.4$) with varying
position angles with respect to the x-axis ($\alpha$).
The parameters of these models are given in Table~1.
\begin{table}
\centering
\begin{tabular}{cccc} \hline\hline
Model & $M_{\rm r}$       & $e$ & $\alpha$ \\ \hline
A     & 0.0               & 0.0 & --       \\
B     & $10^8$            & 0.0 & --       \\
C     & $5\!\times\!10^8$ & 0.0 & --       \\
D     & $10^9$            & 0.0 & --       \\
E     & $10^9$            & 0.4 & 0        \\
F     & $10^9$            & 0.4 & 60       \\
G     & $10^9$            & 0.4 & 90       \\ \hline \\
\multicolumn{4}{c}{Table 1: Model Parameters}
\end{tabular}
\end{table}

The surface density produced by model~A which has no ring is shown in
Figure~1. 
\begin{figure}[t]
\plotfiddle{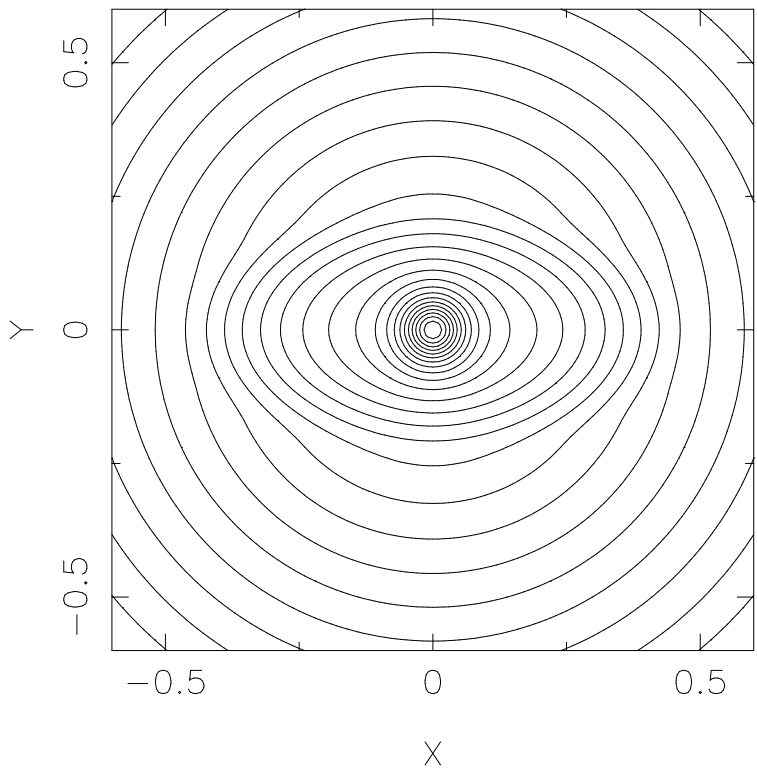}{9cm}{0}{80}{80}{-252}{-162}
\caption{Surface density of model without nuclear ring.  Each contour
represents a change in density of 20\%.  The frame is 12\,Kpc on a side.}
\end{figure}
Because of the small dense central region, the rotation curve of the 
azimuthally averaged mass distribution rises steeply to about 
180\,Km/s at 500\,pc, followed by a more gentle rise to 240\,Km/s at 6\,Kpc.

\section{METHOD}

We determine the dynamical consequences of the nuclear ring on stars and gas
by studying the orbits that are supported by the potential.  The backbone
of an orbital analysis are the periodic orbits, i.e. orbits which
make a closed figure in a frame of reference that rotates with the bar.
Since most orbits in a galactic like potential are not periodic,
one may ask what is the importance of these, rather special orbits.
The answer is that the periodic orbits can be used to characterize the
overall orbital structure of phase space. Each periodic orbit traps a
region of phase space around it, so that nearby orbits within the region
trapped will remain nearby. Trapped orbits generally will have a similar
shape to its parent periodic orbit. 
Such orbits are referred to as being regular, while orbits which are not 
trapped around any periodic orbit, are referred to as being irregular
or ergodic, and are free to move throughout the non-regular regions, 
at least within energy considerations.

It should also be noted that in a rotating frame of reference 
neither the energy nor the angular momentum are conserved.  There is however 
an integral of motion, known as the Jacobi Integral.  This integral of 
motion, which remains constant along any given orbit, can be thought 
of as an effective energy.

To locate the periodic orbits we systematically search the phase
space ${\bf y}_0 = \{x_1,x_2,x_3,\dot{x}_1,\dot{x}_2,\dot{x}_3\}$,
by computing orbits with a given Hamiltonian and starting
on a particular plane $x_i=0$ with sign $\dot{x}_i$.  The Hamiltonian,
given in this case by the Jacobi energy $\Ej$, is used to
eliminate the momentum conjugate to $x_i$, leaving four degrees of
freedom for the initial conditions.  The trajectory is
then computed, in the rotating frame of the bar, until it 
again crosses the plane $x_i=0$ with the same sign as $\dot{x}_i$,
giving phase space coordinates $T({\bf y}_0) = {\bf y}_1$.  
The periodic orbits given by the fixed points of the map 
$T({\bf y}_0) = {\bf y}_0$ are
then computed by an iterative procedure based on the numerically determined 
Jacobian $\nabla\!T$.  For this the least-squares method described by 
Pfenniger and Friedli \markcite{pfe93} (1993) is used. Further details
on finding periodic orbits may also be found in this reference.

Since the Jacobian $\nabla\!T$ is a linear approximation of the
phase space around ${\bf y}_0$, i.e.
\begin{equation}
T({\bf y}_0 + \Delta{\bf y}) = T({\bf y}_0) + \nabla\!T\cdot\Delta{\bf y}
	  + O(\|\Delta{\bf y}\|^2),
\end{equation}
its eigenvalues provide information on the stability of the periodic orbit.
For convinience, two stability coefficients based on these eigenvalues
are defined, such that the orbit is stable if both are real and 
less than or equal to two in absolute value (Pfenniger \markcite{pfe84} 1984).
For planar orbits, the eigenvectors have fixed directions, and the
stability coefficients my be identified with the horizontal ($b_{\rm H}$)
and vertical ($b_{\rm V}$) stability of the orbit.  

The orbits are computed using a 13 stage embedded Runge-Kutta
formula of order 8/7 with adaptive step size (Prince \& Dormand 
\markcite{pri81} 1981).  The trajectories are computed with a 
relative accuracy of $10^{-14}$.
The program uses the Parallel Virtual Machine (PVM) software
package to distribute the computation over multiple processors.
The parallelization algorithm features dynamical load balancing and 
is robust against the lose of machines.  Both of these features are
accomplished by the use of a circular queue implemented by a link list.
Each energy to be searched is inserted into the queue.  Each slave
process searches one energy at a time, requesting an additional
energy from the master process as needed.  As the master process
gives out an energy to a slave the pointer to the head of the queue
is advanced. Because of its circular nature this effectively moves the 
energy to the back of the queue.  As the number of energies in the
queue becomes smaller than the number of slaves the same energy
may be given out more than once.  An energy is only removed from
the queue when a slave has successfully completed the search at
that energy and returned to the master any results.  The full search
is complete when the queue is empty.  For this study eight Dec-Alpha 
workstations were used.

We start by first locating all the low order periodic orbits in the
disk plane and determine the sections along the families which are
vertically unstable.  For each of these vertical instability strips
we search for bifurcating vertical orbits.  Once a vertical
orbit has been located, its family is followed by predicting the
phase space position of subsequent orbits using a fourth order
backward divided differences polynomial with variable step size.

\section{RESULTS OF ORBITAL ANALYSIS}

\subsection{Without a Nuclear Ring}

We start by locating the low-order planar periodic families inside
corotation when no ring is present. The search is restricted to simple
periodic orbits with initial conditions of $x=\dot{y}=z=\dot{z}=0$.
The characteristic diagram is shown in Figure~2.
\begin{figure}[tp]
\plotfiddle{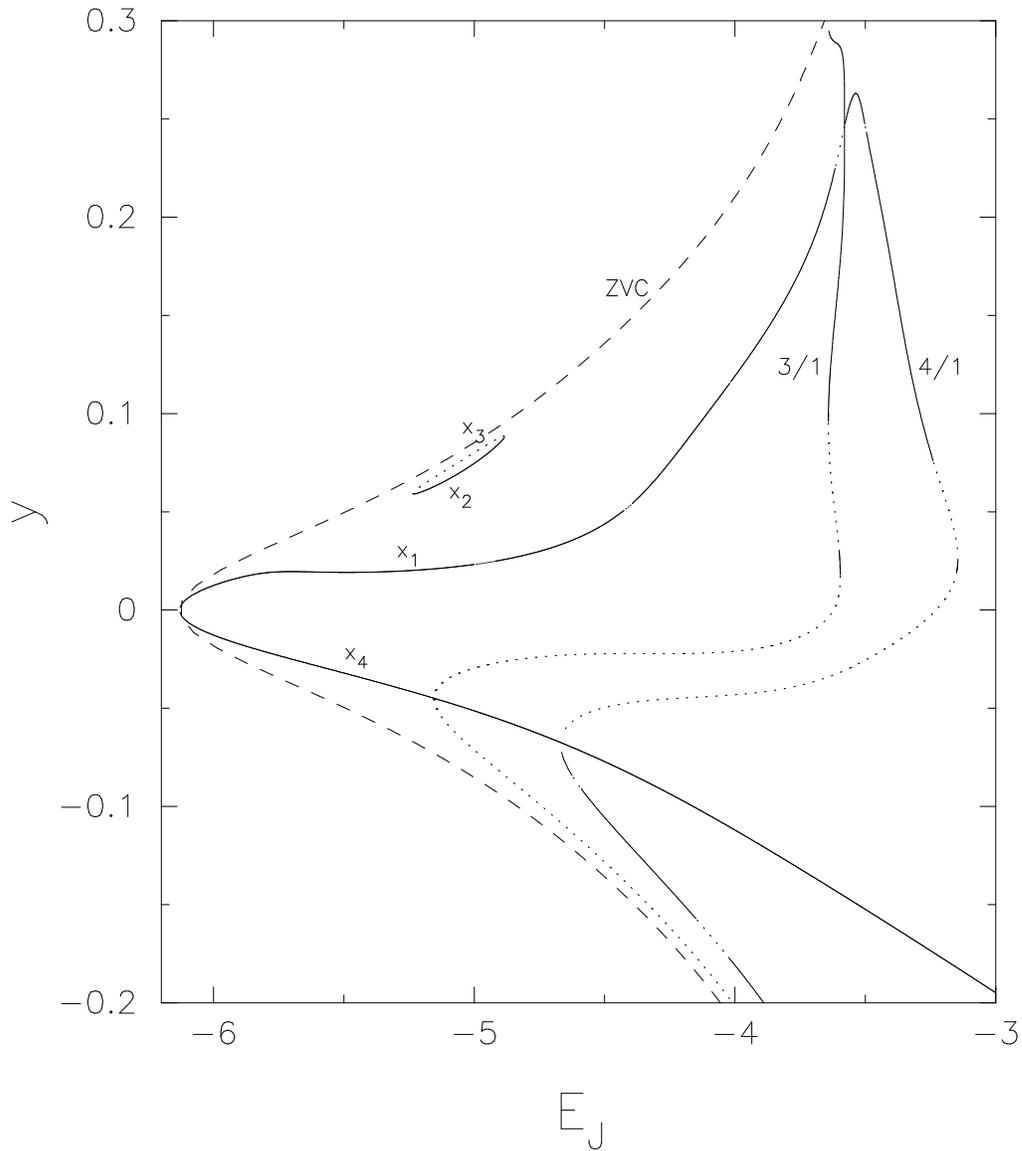}{14cm}{0}{70}{70}{-234}{-60}
\caption{Characteristic diagram for planar simple periodic orbits 
inside corotation when no ring is present. The main families have 
been labeled: direct ($x_1$) and retrograde ($x_4$) orbits elongated 
along the bar; ($x_2$) and ($x_3$) direct orbits elongated perpendicular
to the bar. The dashed curve labeled ZVC is the zero 
velocity curve.  Stable sections are represented by solid lines while
unstable are dotted.}
\end{figure}
This figure shows the Jacobi energy, $\Ej$, and the value of the
crossing point $y$, perpendicular to the bar, for each periodic orbit
located.  The zero velocity curve which delineates the accessible 
region in the plane is denoted by the symbol ZVC. 
While each point in this diagram represents a single periodic orbit, a quick
glance shows that it is not covered uniformly by points, but that
the periodic orbits define curves in this plane. Hence, the orbits are not 
unrelated, but fall into families, with each family being defined by
its characteristic curve.  Sections of the characteristics where 
the orbits are stable are represented with a solid line, while unstable 
sections are broken.

The main families in this diagram have been labeled using the
notation of Contopoulos and Papayannopoulos (1980).  The $x_1$ family are 
orbits elongated along the bar and rotate about the center in the same sense as
the bar (direct). In the frame of reference of the bar,
they also move radially in and out, twice for once around the center (2/1).
It is the population of these orbits that predominantly
defines the bar, giving it it's shape and structure.

The x-extent (along the bar) of the $x_1$ family increases 
monotonically with increasing energy from the center out to the 3/1 
bifurcation, while the y-extent has a local maximum at $\Ej=-5.70$.
Due to the moderately weak bar, the $x_1$ family orbits have no 
loops at the ends.
The eccentricity also increases monotonically from the center, but
only until $\Ej=-4.80$ where it has reached a
maximum value of 0.97, and the x-extent of the orbits
is about 1.1\,Kpc.  Further out, the eccentricity of $x_1$ orbits stays
almost constant, and then declines sharply just before the corotation
radius.  At an energy of $\Ej=-3.54$ the $x_1$ sequence 
turns downward and changes to 4/1.  In principle, this section of the
characteristic is no longer the $x_1$ orbits as designated by the 
notation of Contopoulos, which continues to higher energies following
a large gap.  The 4/1 sequence changes from direct to retrograde at
$\Ej=-3.23$, and then 3-periodic at $\Ej=-3.63$, before 
meeting the retrograde family $x_4$.  This sequence is simpler than
the one studied by Pfenniger \markcite{pfe84} (1984) because of the lack 
of an elbow feature, but is very similar to that described 
by Athanassoula etal \markcite{ath83} (1983).

The family $x_2$ are direct orbits elongated perpendicular
to the bar and their presence is indicative of a ILR(s) in the
nonlinear regime (van Albada \& Sanders \markcite{van82} 1982). 
The characteristic covers a range in energy from $\Ej=-5.24
\mbox{ to }-4.88$ along which the y-extent of the orbits increases
monotonically from 0.59 to 0.88\,Kpc, while the x-extent goes
from 0.33\,Kpc to a local maximum of 0.46\,Kpc at $\Ej=-4.94$. 
These four points may be used to define the position of a
double radial ILR along each axis. As usually, the $x_2$
orbits are more rounded than the $x_1$ orbits, with the eccentricity
increasing towards higher energies.
The $x_3$ family are also direct orbits elongated perpendicular
to the bar, but are everywhere horizontally unstable.

The $x_4$ family are also 2/1 orbits, but travel in a retrograde
sense to the bar and are slightly elongated perpendicular
to the bar.  For consistency with a bar, this family can only
be sparsely populated, as disks with a significant fraction of
retrograde orbits are stable to the formation of bars.  Likewise,
the population of this family during the later stages of evolution
can play a role in the dissolution of a bar.

While the $x_2$, $x_3$, and $x_4$ families are vertically stable 
along their full characteristics, the $x_1$ family has several
sections of vertical instability.
We are primarily interested in the two vertically unstable strips at
the lowest energies. They can be seen more clearly in the top
frame of Figure~3 where they are labeled $S_1$ and $S_2$.
\begin{figure}[tp]
\plotfiddle{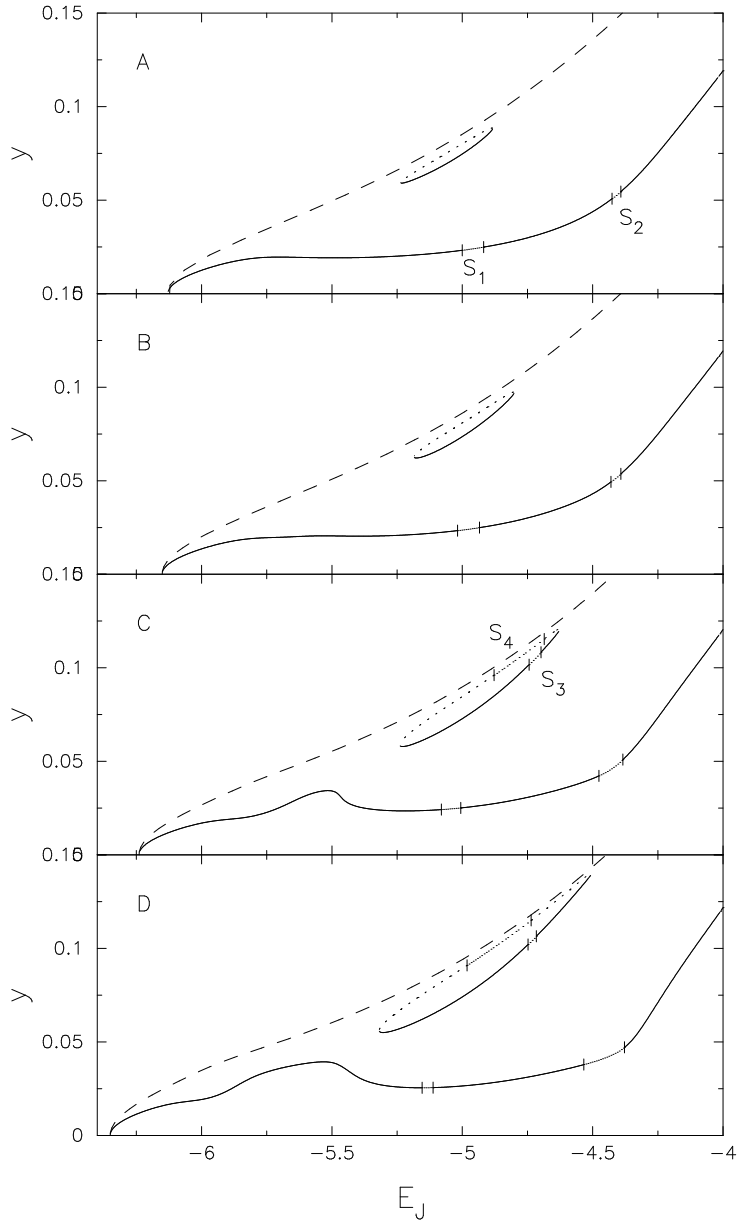}{14cm}{0}{70}{70}{-216}{-30}
\caption{Characteristic diagrams of the $x_1$, $x_2$, and $x_3$  families
for models with ring mass: (A) $0.0$; (B) $10^8$; (C) $5\!\times\!10^8$;
and (D) $10^9\,M_{\sun}$.  Stable sections of the characteristics
are represented by solid lines while unstable are dotted. Four vertical
instability strips are marked. The dashed curve is the zero velocity curve.}
\end{figure}
In the top left frame of Figure~4 are the stability coefficients for
this family. Recalling that a value of $|b|>2$ indicates instability, the
diagram shows that the two strips are vertically unstable.
\begin{figure}[tp]
\plotfiddle{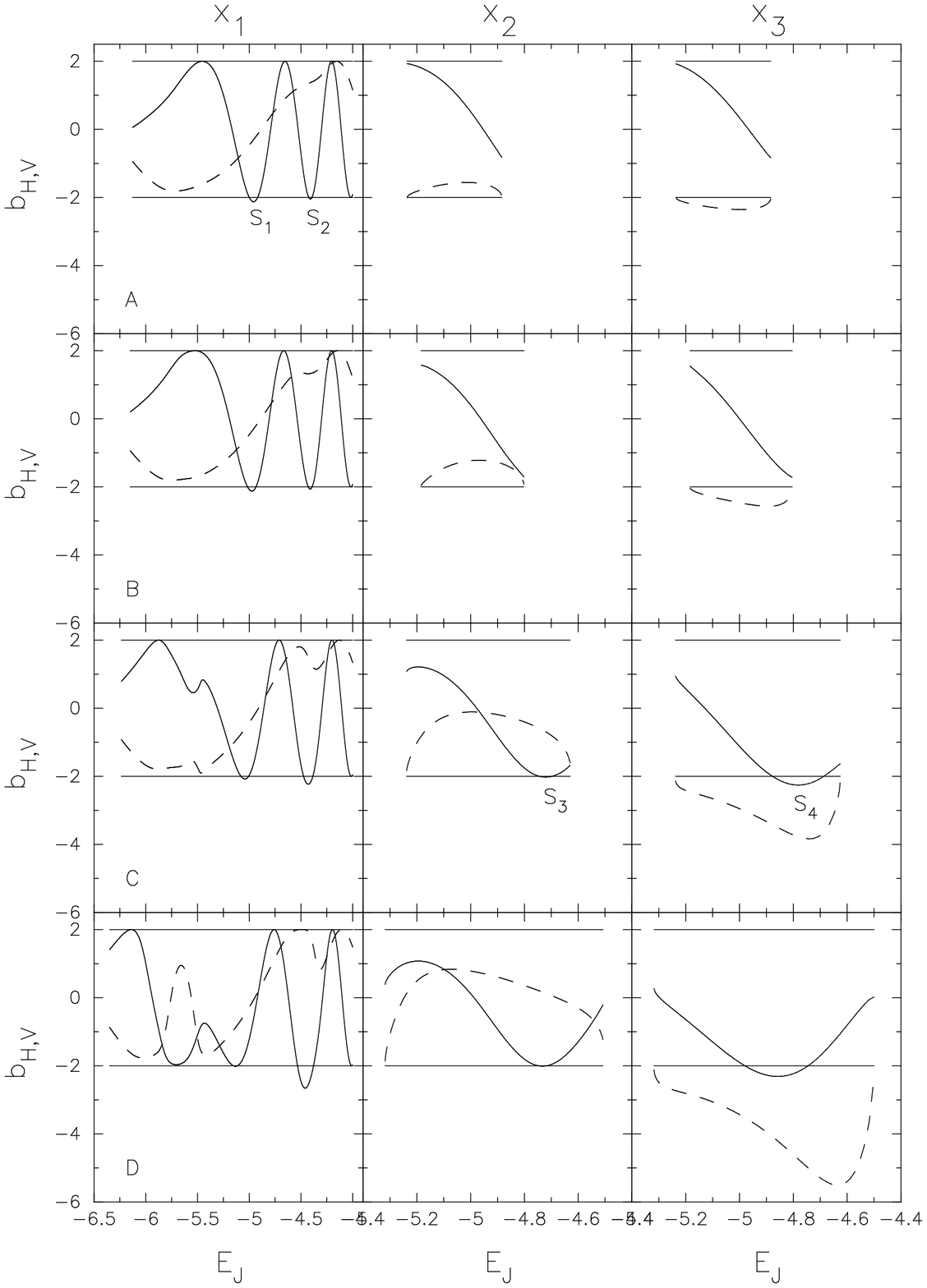}{14cm}{0}{70}{70}{-216}{-30}
\caption{Stability diagram showing horizontal coefficient $b_{\rm H}$
(dashed) and vertical coefficient $b_{\rm V}$ (solid) for the planar
periodic orbits of the four models. The stability of the $x_1$, $x_2$,
and $x_3$ is shown from left to right.  The four vertical instability 
strips are indicated.}
\end{figure}

Both of these strips, $S_1$ and $S_2$, have two vertical families of orbits 
bifurcating from them, one in $z$ and the other in $\dot{z}$.  
Their characteristics are shown in the top frames of Figures~5 and 6.
\begin{figure}[tp]
\plotfiddle{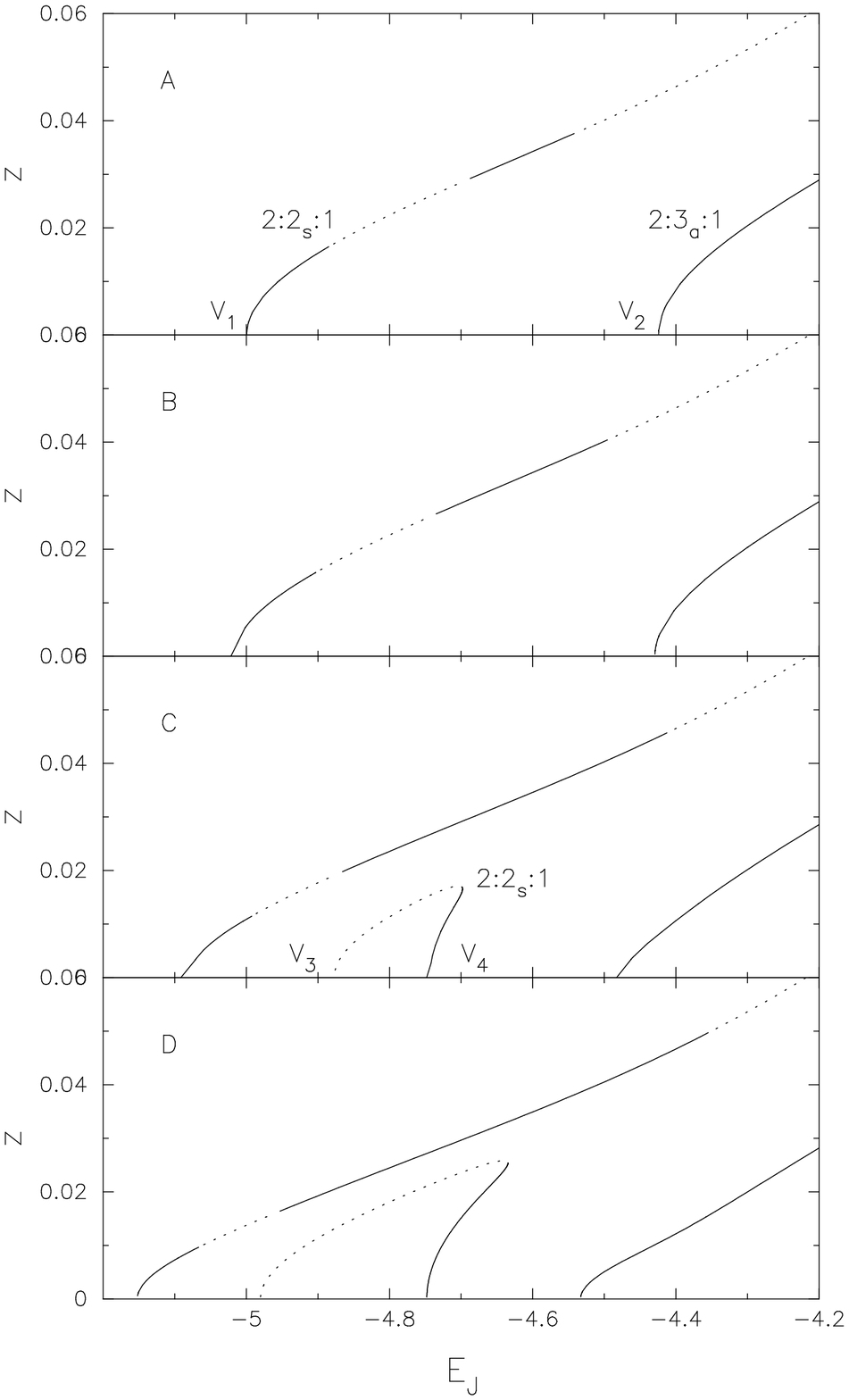}{13cm}{0}{70}{70}{-216}{-35}
\caption{Characteristic diagram of vertical orbits bifurcating from
the planar instability strips in $z$.  Stable sections of the
characteristic are shown as a solid line while unstable sections
are dotted.}
\end{figure}
\begin{figure}[tp]
\plotfiddle{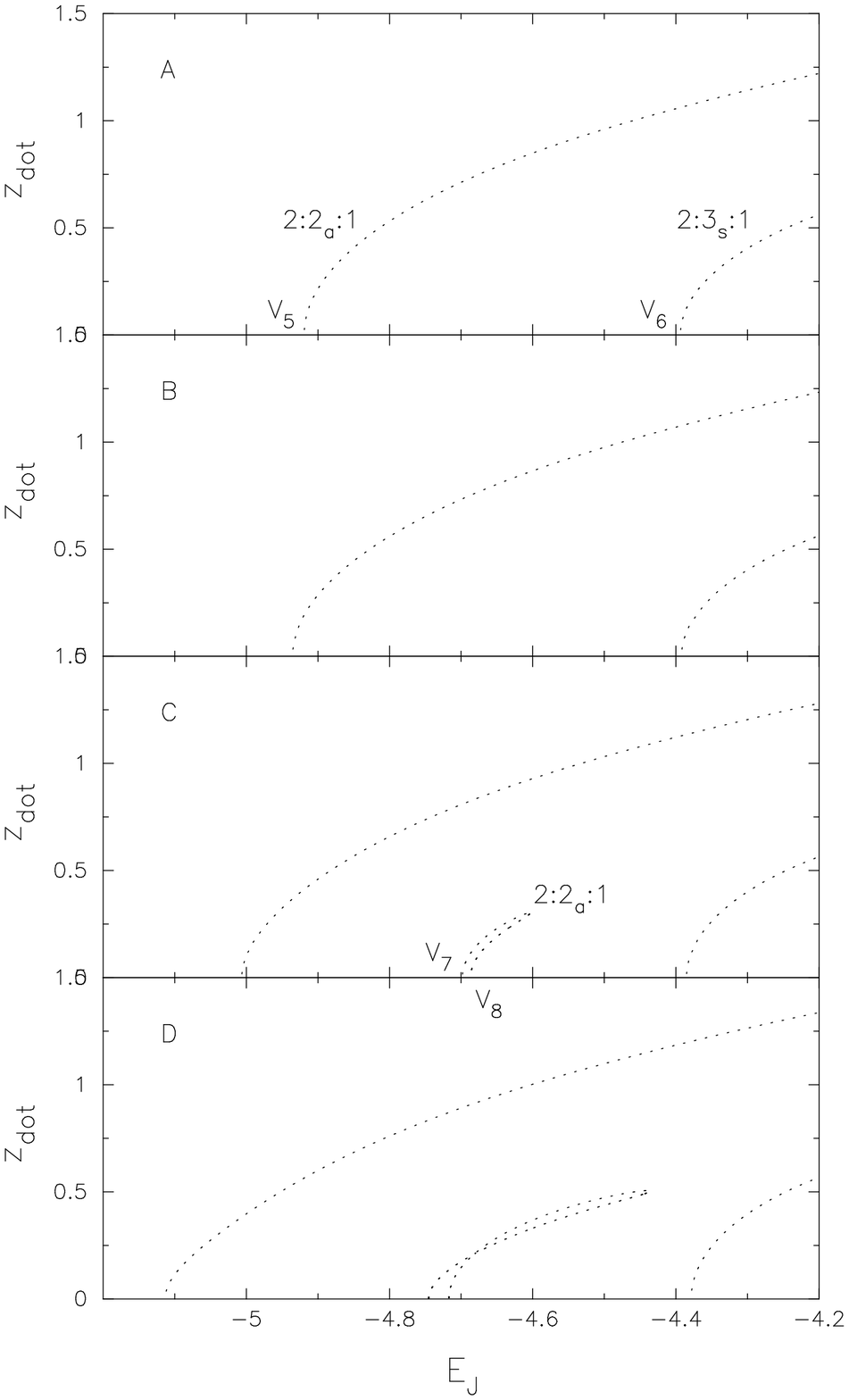}{13cm}{0}{70}{70}{-216}{-35}
\caption{Characteristic diagram of vertical orbits bifurcating from
the planar instability strips in $\dot{z}$.  All of the orbits
depicted here are unstable.}
\end{figure}
We classify the 3D orbits using the notation of Sellwood and
Wilkinson \markcite{sel93} (1993).  In this notation $m\!:\!n\!:\!l$
implies $m$ radial oscillations in the $(x,y)$ plane and $n$ vertical
oscillations in $z$ as the orbit closes $l$ rotations about the center.
An `s' or `a' subscript is added to indicate whether the orbit is
symmetric or anti-symmetric about the $(y,z)$ plane.

The vertical bifurcation $V_1$ in $z$ leads to a family of 
$2\!:\!2_{\rm s}\!:\!1$ orbits (BAN, i.e. banana-shaped orbits)
which are stable from the point of bifurcation at $\Ej=-5.00$ up
to $\Ej=-4.88$, then again between $\Ej=-4.69$ and
$\Ej=-4.54$, after which it remains unstable over the region
of interest for this study. The unstable sections shown in Figure~5
are {\it complex} unstable (see Hasan etal \markcite{has93} 1993 and 
refs. therein for discussion of this instability). The vertical bifurcation
$V_2$ in $z$ at $\Ej=-4.42$ leads to a family of $2\!:\!3_{\rm a}\!:\!1$ 
orbits which are stable over the entire region of interest.  The 
vertical bifurcations $V_5$ at $\Ej=-4.92$ and $V_6$ at $\Ej=-4.39$ in 
$\dot{z}$ lead to $2\!:\!2_{\rm a}\!:\!1$ (ABAN, i.e. anti-banana orbits) and 
$2\!:\!3_{\rm s}\!:\!1$ orbits, respectively, which are unstable everywhere. 
This situation, where the $z$ and $\dot{z}$ bifurcations lead
to symmetric and anti-symmetric pairs of orbits, is expected because of
the symmetry of the phase space.  However, unlike the $2\!:\!2\!:\!1$ orbits
which are either symmetrical or anti-symmetrical about both the $(x,z)$ and
$(y,z)$ planes, the $2\!:\!3\!:\!1$ orbits are symmetrical with respect
to one plane and anti-symmetrical with respect to the other.
The $2\!:\!3\!:\!1$ orbits are shown in projection and in a ``tube'' view
in Figures~7 and 8.
\begin{figure}[tp]
\plotfiddle{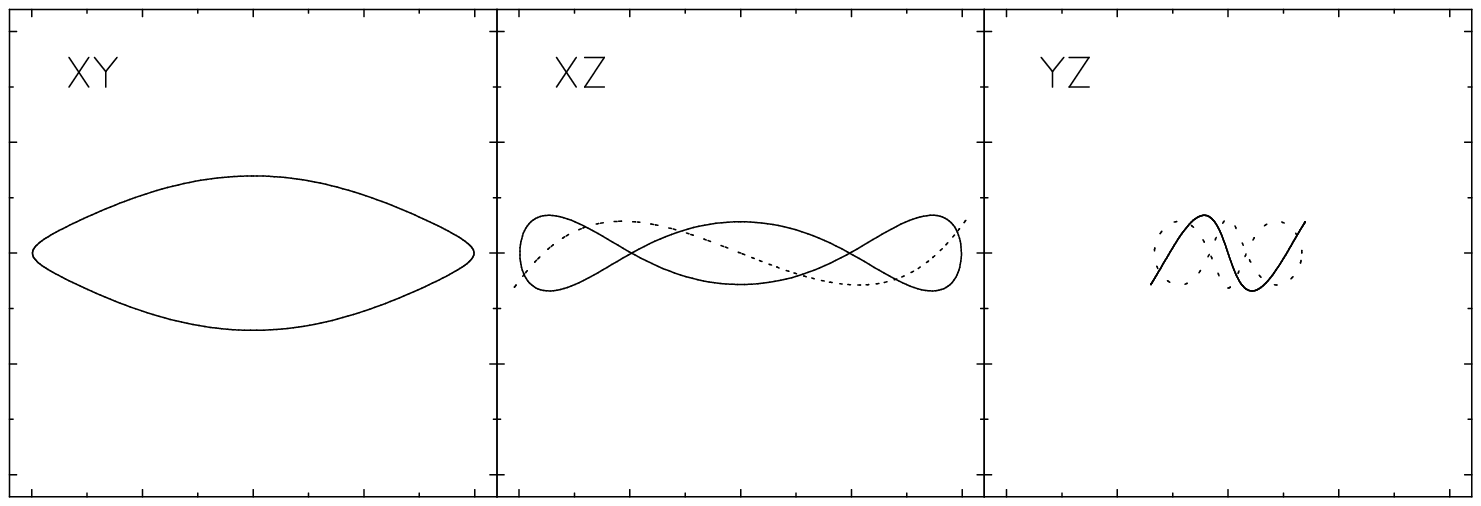}{6cm}{0}{110}{110}{-320}{-340}
\caption{ Projections onto the three planes of the $2\!:\!3\!:\!1$
symmetric (solid) and anti-symmetric (dashed) 3D orbits.}
\end{figure}
\begin{figure}[tp]
\plotfiddle{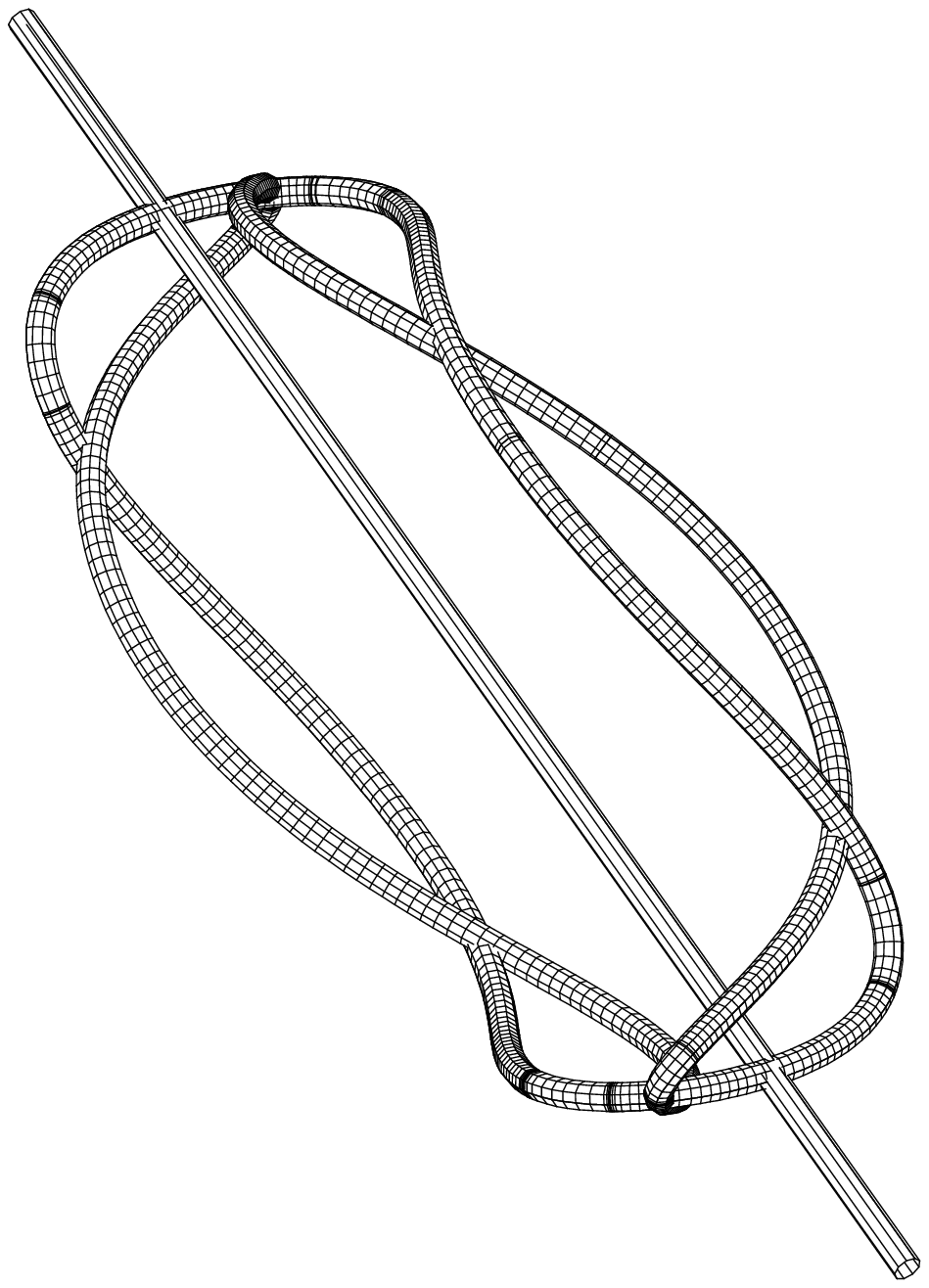}{9cm}{-90}{100}{100}{-400}{460}
\caption{Tube view of the $2\!:\!3\!:\!1$ orbits. The x-axis is
also indicated.}
\end{figure}

\subsection{With a Circular Nuclear Ring}

The previous section described the basic orbital structure of model~A which
contains no nuclear ring.  We will now examine the change in the orbital
structure when an increasingly massive circular ring is introduced. 
Position of the ring on the x-axis was chosen as to be between the original
ILRs of the Myamoto-Nagai model (0.33 kpc and 0.46 kpc, 
from extension of $x_2$ orbits). A circular ring crosses the y-axis interior 
to the inner ILR. Because the ring is smeared in the radial direction, it
actually fills up all the space between the ILRs.

The strongest effect observed with the growth of the ring (Fig.~3) is that the
extent in energy covered by the $x_2$ and $x_3$ orbits increases
as the ring mass increases.  Previous studies
have shown such behavior when the bar is weakened (e.g. Contopoulos
\& Papayannopoulos \markcite{con80} 1980). In the present study, where
the $x_1$ family of orbits exists also interior to the ring, the bar is
further weakened by a developing and widening shoulder (``bump'')
around $\Ej\approx-5.50$, in the vicinity of the ring. This ``bump'' is 
responsible for $x_1$ orbits
becoming rounder near the inner ILR. A comparison of the $x_1$
and $x_2$ orbits in the models A and D can be seen in Figure~9.
\begin{figure}[tp]
\plotfiddle{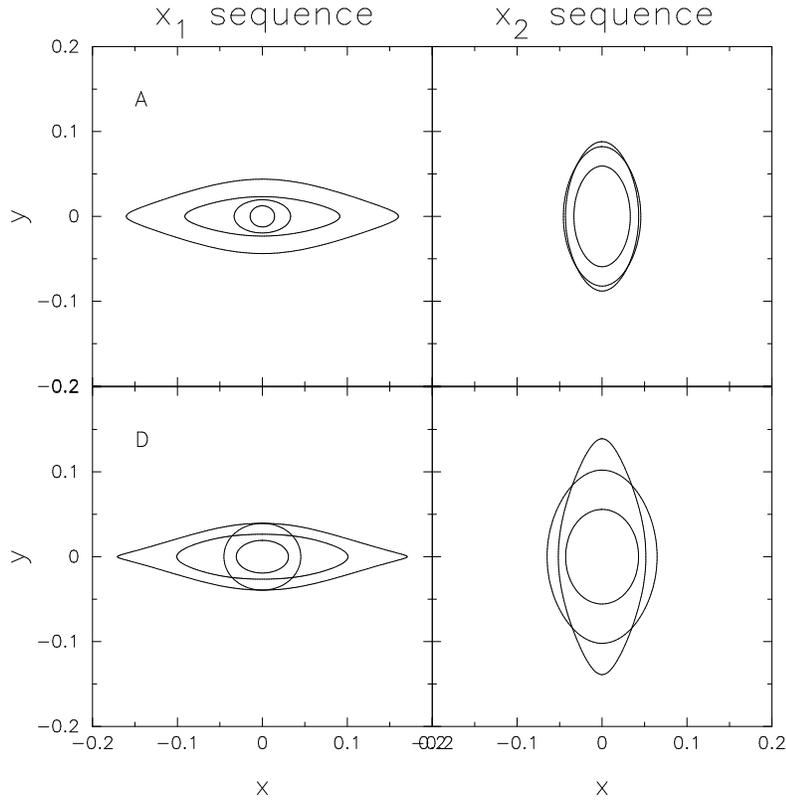}{10cm}{0}{60}{60}{-185}{-80}
\caption{Comparison of $x_1$ and $x_2$ orbits between the models A and D.
For the $x_1$ orbits $\Ej=-6.0,-5.0,-4.5$, and the one with maximum
y-extent in the ``bump'' is shown.  For the $x_2$ orbits the minimum 
and maximum energy, along with the orbit of maximum x-extent is shown.}
\end{figure}
The development of the ``bump'' produces a significant local maximum in 
the y-extent of the $x_1$ orbits, while the x-extent continues to increase 
monotonically with increasing energy.  This causes the $x_1$ orbits in the 
presence of a massive ring, e.g. in model~D, between 
$\Ej=-5.53$ and -4.50, or equivalently of 0.45 to 1.7\,Kpc x-extent,
to intersect with $x_1$ orbits of lower energy. Similarly the local
maximum in x-extent along the $x_2$ sequence causes the orbits with
$\Ej=-4.75$ to -4.50, or 0.65 to 0.52\,Kpc x-extent, to intersect
with lower energy $x_2$ orbits. Due to the $x_2$ orbits being rounder than
the $x_1$ orbits, they are less distorted and start to intersect at larger
ring masses.

As the ring mass is increased, the instability strip $S_1$ 
shrinks while at the same time shifting to lower energies.  In contrast
the instability strip $S_2$ grows in size toward lower energies.
This can be seen more clearly in
Figure~4 where the increasing ring suppresses $S_1$ while enhancing $S_2$.
For the two higher mass models C and D, vertical instability strips 
$S_3$ and $S_4$ form, respectively, in the $x_2$ and $x_3$ sequences, 
which is rather unusual, as the $x_2$ orbits are typically stable.
The only case of instability strips on $x_2$ orbits connected by vertical
bifurcations was reported by Udry (1991) for slowly rotating triaxial
systems. There is only remote similarity between the latter and our case,
in that the instability appears in highly flattened systems with $c/a<0.3$.  
However, contrary to our case, Udry's bifurcations connect two instability
strips on the same $x_2$ sequence.

From Figures~5 and 6, we see that the vertical bifurcation points
$V_1$, $V_2$, and $V_5$ on the $x_1$ sequence shift to lower energies
as the ring increases, while $V_6$ barely moves.  This is in agreement
with the evolution previously noted for the instability strips, since the
vertical bifurcations occur at the edges of each strip.
Of particular significance are the changes in the characteristic which
bifurcates from $V_1$, where the extent of stable orbits increases
significantly.
We also see the creation of a $2\!:\!2_{\rm s}\!:\!1$ (BAN) orbital
family, forming a loop between the bifurcation points $V_3$ and $V_4$,
respectively on $x_3$ and $x_2$, which grows larger with increasing ring
mass. Half the loop, up to the maximum in $z$, is stable while the other
half is unstable.  A family of $2\!:\!2_{\rm a}\!:\!1$ (ABAN) orbits
also connects $S_3$ and $S_4$ through a bifurcation in $\dot{z}$ at $V_7$
and $V_8$, but is unstable everywhere, with a section on the high energy
side being complex unstable. These non-planar orbits are elongated
perpendicular to the bar and connect the planar $x_2$ and $x_3$ families.

\subsection{With an Elliptical Nuclear Ring}

To examine the effect of a non-circular ring, we have run models with
the maximum ring mass (as in model~D), a small eccentricity ($e=0.4$), 
and at varying position angles with respect to the bar. We find that for
rings which are either aligned with ($\alpha=0\arcdeg$) or perpendicular
($\alpha=90\arcdeg$) to the bar, the differences with the circular ring
are small.  The planar characteristics of model~D, which has a circular 
ring, and model~E, with the ring major-axis aligned with the bar, differs 
only in that the ``bump'' of model~E is slightly smoother.  For model~G, with
the ring major-axis aligned perpendicular to the bar, the break
in the $x_1$ characteristic at the high energy side of the ``bump''
is sharper and the $x_2/x_3$ loop extends to a slightly lower value of $\Ej$.
A qualitatively different behavior of $x_1$ and $x_2$ orbits is obtained
when the ring major-axis is oblique to the bar axes. 

Unlike stellar orbits, gas orbits are subjected to dissipational processes.
This has the effect of broadening the range of orbital frequencies which
respond to a resonance.  In contrast to the stellar orbits which may abruptly
change orientation by $90\arcdeg$ at an ILR, the gas orbits
make a slower transition.  Depending on the details of the potential
and pattern speed, the gas may settle in the ILR region on orbits 
which are neither aligned nor perpendicular to the bar, but instead at
some intermediate angle.

This configuration is the one typically observed for nuclear rings and 
also appears as a long-lived transient in our numerical simulations
(Knapen etal. 1995a).  We investigate this by making the ring major-axis
oblique to the bar axes.  In this case, we observe that the major-axes of 
the $x_1$ orbits become phase shifted, with the position angle of their
axes a function of radius. This is shown in Figure~10, where we have
plotted some representative $x_1$ orbits of model~F.
\begin{figure}[tp]
\plotfiddle{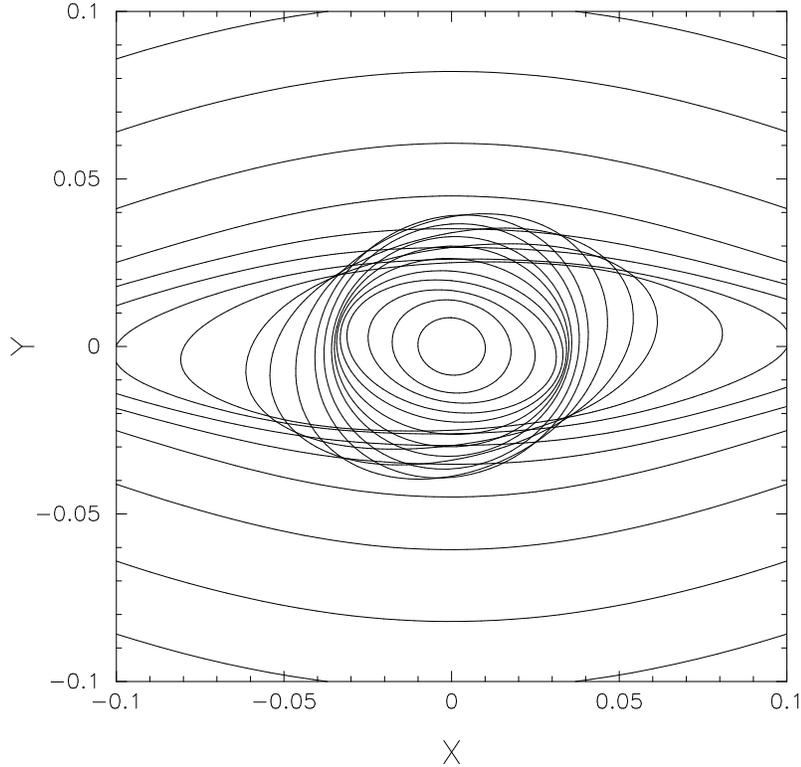}{10cm}{0}{60}{60}{-188}{-80}
\caption{Twisting of $x_1$ orbits from model~F with oblique elliptical
ring. The ring with an ellipticity of $e=0.4$ and semi-major axis
$R_0=0.04$ is offset to the bar by $\alpha=60\arcdeg$ in the leading
direction. The frame is 2\,Kpc on a side.}
\end{figure}
The change in eccentricity and position angle of
the $x_1$ orbits as a function of their semi-major axis is shown in Figure~11.
\begin{figure}[t]
\plotfiddle{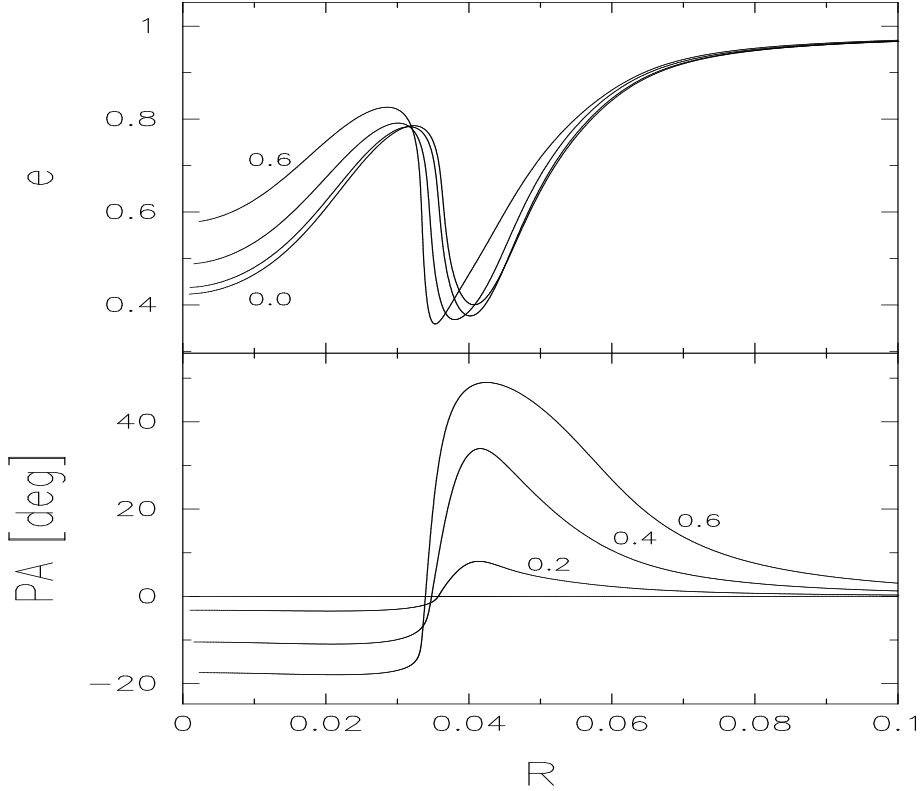}{10cm}{0}{80}{50}{-245}{-35}
\caption{Eccentricity and position angle of $x_1$ orbits as a function
of their semi-major axis.  The models are for a $10^9\,M_{\sun}$ ring
leading the stellar bar by $60\arcdeg$ and with eccentricities of 
0.0, 0.2, 0.4, and 0.6.}
\end{figure}
The models are for a $10^9\,M_{\sun}$ ring
leading the stellar bar by $60\arcdeg$ and with eccentricities of
0.0, 0.2, 0.4, and 0.6. The eccentricity of the nuclear ring does show a strong effect on the position angle of the $x_1$ orbits, both outside and inside the
ring (Fig.~11). In addition, the eccentricity of the $x_1$ orbits interior to
the ring is increased with the eccentricity of the ring itself.

The $e=0.4$ ring (model~F) differs little
from the circular ring ($e=0.0$, model~D) in the trend of orbit
eccentricity with semi-major axis except for a small shift.  The 
eccentricity of model~F reaches a maximum of  
$e=0.79$ at  $R=0.030$ or $\Ej=-6.03$, followed by a minimum of
$e=0.37$ at $R=0.038$ or $\Ej=-5.72$. However, the change in the position
angle with $R$ is rather dramatic, starting at large radii 
with a slowly growing phase shift which reaches a maximum of  $34\arcdeg$
in the leading direction at $R=0.042$ or $\Ej=-5.61$, followed by a fast
decline to $0\arcdeg$ at $R=0.035$ or $\Ej=-5.86$, and then continuing to 
$-11\arcdeg$ in the trailing direction at $R=0.021$ or $\Ej=-6.16$, after
which, remarkably, it remains offset in the trailing direction deep into
the bulge. As can been seen from the Figure 11, the minimum in eccentricity
is slightly offset from the maximum position angle in radius.

This behavior of $x_1$ orbits can be understood within the framework of
forced oscillations. The new thing here is that the orbit is responding
to two external forces, the stellar bar and an oblique ring, which have 
the same driving frequency but are phase-shifted. A straightforward 
application of the epicyclic approximation (see Appendix A) shows that the main 
periodic orbits respond by twisting the orientation of their major axes with
respect to the stellar bar. Even in the case when
the ring is more or less circular, but the mass distribution has a
quadrupole moment, should the effect remain qualitatively the same.
Further technical details are provided in Appendix A.

\section{SUMMARY AND CONCLUSIONS}

We have performed an orbital analysis in a 3D galactic potential perturbed
by a growing nuclear ring. The galaxy consisted of a disk, bulge and a bar.
We have studied the effects of circular and oval rings positioned
in the preexisting double radial ILR region of this system. 

Different orientations of the oval ring with respect to the stellar
bar have been tested as well: aligned, perpendicular, and leading by
$60\arcdeg$ in the direction of galactic rotation. In all cases, both 
the ring's semi-major axis and the initial outer radial ILR were 
smaller than the semi-minor axis of the bar by a factor of $\sim3$. 
The ring's mass was always negligible compared to the overall mass 
of the system, but is responsible for a local distortion of the 
axisymmetric rotation curve.

We find that the major periodic orbits within the corotation radius
are profoundly affected by the perturbation of the ring. For both circular
and oval rings we observe a
substantial increase in the phase space allowed by the $x_2$ orbits.
If these orbits would be populated, the bar would be weakened.
The biggest change comes from the outer radial (and vertical) ILR
moving further out to higher energies and increasing its x- and y-extents.
The inner ILR moves inwards only slightly. Hence the growing mass in the ring 
affects mostly the orbits outside the ring (this can be seen even more
explicitly in the case of the oblique ring of model F). The only effect the
ring has on the inner $x_1$ orbits is the appearance of a shoulder (bump)
interior to the inner ILR in the characteristic diagram (Fig. 3). This 
reflects the fact that the $x_1$ orbits in this region become more rounded.
Also, the vertical instability strips of the $x_1$ orbits, $S_1$ and $S_2$,
are moving closer to the radial ILRs, i.e. to lower energies. Whereas $S_1$  
becomes narrower, $S_2$ widens substantially towards lower energies (Fig.3).
The $x_2$ orbits acquire vertical instability strips as well, when the
ring becomes massive. This has some similarity to the slowly rotating
triaxial models of Udry (1991).

Weakening of the bar, when increasing the central mass concentration in a
galaxy by adding a massive BH and/or bulge, was reported in all relevant 
studies (Pfenniger and Norman \markcite{pfe90} 1990; Hasan and Norman 
\markcite{has90} 1990; Hasan, Pfenniger \& Norman \markcite{93} 1993).
It was associated with the appearance of a single radial (and vertical)
ILR due to the altered rotation curve.  This ILR moved out rapidly with
the growing central mass $M_{\rm c}$, specifically $R_{\rm ILR}\propto
M_{\rm c}^{2.8}$ (e.g. Pfenniger \markcite{pfe96} 1996). Under these 
circumstances, chaotic orbits dominate the space between the center 
and the ILR, which quickly approaches the semi-minor axis of the bar, causing 
the dissolution of the stellar bar in a secular process. In addition, higher 
order radial and vertical ILRs contribute to this effect by widening 
the resonance zone.  The growing central mass repells the $x_1$ family
of orbits from the central region, i.e. within the ILR, and by doing so
it also affects the motion out of the xy-plane. This happens because of two
reasons (Hasan, Pfenniger and Norman \markcite{has93} 1993). First, the 
vertical bifurcation points which mark the origin of 3D orbits, as described
in Section 4, move away from the center as the central mass grows.
Second, the stability of the out-of-the-plane orbits is affected, with a
region of instability appearing in the innermost family of orbits, which
corresponds to our $V_1$ (2:2$_{\rm s}$:1) orbits.

The model presented here differs from the above scenario mainly in two 
aspects: the preexisting {\it double} radial (and a single vertical) ILRs, 
and the differing symmetry in the perturbing potential, which is far from
being spherically-symmetric. Consequently, the $x_2$ orbits continue to be 
limited only to the space between the radial ILRs, whereas the $x_1$ orbits 
still support the bar between the center and the inner ILR, and between the 
outer ILR and corotation. This is true irrespective of the ring's mass, 
shape or inclination to the bar. The radius of the outer ILR grows 
approximately {\it linearly} with the mass in the ring, slower than in 
the case of the central mass increase.  In addition, an oval ring, oblique
to the bar, further degrades the symmetry in the galactic plane, which is
otherwise dominated by the $m=2$ mode due to the stellar bar. As a result
the $x_1$ orbits within $\sim2-3$ semimajor axes outside the ring, and 
at the position of the ring, are significantly distorted. We note, that
such shapes and inclinations of a ring are not merely of a theoretical
possibility, but are in fact supported by observations of molecular gas
distributions in nuclear starburst galaxies.

The above differences, between the dynamical effects of a central mass
and of a ring, create interesting possibilities for bar evolution in 
response to the growing mass in the ring. In both cases the stellar bar 
should weaken as the supporting $x_1$ family of orbits are destroyed
or depopulated. However, the ring has a relatively small effect
on the interior orbits, and so we expect that the dissolution of the bar
should mainly proceed at and exterior to the ILR resonance region. 
We speculate, that the observed difference between stellar bars in 
early and late-type disk galaxies (Elmegreen and Elmegreen 1989) may have 
its origin in the secular dissolution of the $x_1$ family of orbits 
beyond the ILR(s).

Another example which is relevant to this discussion and was studied
by us in detail using high-resolution NIR imaging is M100 (=NGC~4321),
a nuclear starburst galaxy of intermediate type SABbc (Knapen etal. 
\markcite{kna95a}\markcite{kna95b} 1995a,b).
This galaxy exhibits all the virtues of a double ILR with a nuclear 
ring. It possesses a 6 kpc semimajor axis stellar bar, bissected by a weakly
oval NIR nuclear ring whose inner boundary, at the inferred position of the
inner ILR, is clearly delineated by an incomplete ring of star formation.
If NIR isophote ellipse fitting provides a reliable measure of the strength of
the barred potential, we measure the maximal ellipticity exterior to the
ring to be approximately 20\% lower than the maximal ellipticity interior
to the ring. It is very interesting, that its NIR isophote twist is similar
to that of Fig. 10 (see also Fig. 11).
Two more examples supporting our conjecture that nuclear rings
weaken mainly the large-scale bars are NGC~4274 and NGC~4643 
(Shaw etal. \markcite{sha95} 1995), whose nuclear barlike features have
higher ellipticities than their primary bars.

A crude estimate from our model shows that the linearly growing radius of the
outer radial ILR will reach the size of the semi-minor axis of the bar
when the mass
in the ring is approaching $\sim10^{10}\,M_{\sun}$, a value which is
comparable to typical masses of galactic bulges. Evidently, the stellar bar
cannot extend to the corotation in this case. Of course, the above model
lacks self-consistency in the sense that the orbit analysis is performed in the
static potentials of a galaxy and stellar bar.
In reality, the stellar bar potential would adjust itself to the loss of 
the main supporting orbits and the dissolution process would be accelerated,
lowering the mass required for bar destruction between the ILRs and the
corotation.

How relevant is the above analysis for the gas evolution in the barred region?
Periodic orbits do not exactly correspond to the gas orbits in a galactic
potential, mainly due to our neglect of pressure forces. However, by virtue
of being the most stable orbits, they can trap gas if the latter is only
weakly dissipative. We find that the $x_1$ periodic orbits 
of different Jacobi energy in our models begin to intersect in the  
vicinity of a ring when its mass provides a sufficient perturbation to the 
background potential. The $x_2$ orbits start to intersect later-on,
when the ring becomes even more massive. This difference in the behavior
of $x_1$ and $x_2$ orbits happens because the latter are more rounded than the 
$x_1$ orbits (see Section 4.1) and must be perturbed stronger in order to
overlap. We estimate roughly that this occurs
when the mass in the ring approaches $\sim10\%-
20\%$ of the mass interior to the ring, and so the self-gravity in the gas
cannot be neglected. The above estimate agrees favorably with the 
results of $N$-body simulations (Wada and Habe \markcite{wad92} 1992; 
Heller and Shlosman \markcite{hel94} 1994; Knapen etal. \markcite{kna95a}
1995a). Such intersecting orbits will be quickly depopulated
by the gas in a steady state flow which can only move to orbits deeper
in the potential well. 

An even stronger effect on the gas is expected in the model with an 
oblique ring (model F). Here the $x_1$ orbits experience a large-angle
rotation with respect to the bar due to the double forcing from the bar
and the ring. Such rotations would almost certainly lead to the intersection
between $x_1$ orbits, which exterior to the ring, have fairly eccentric shapes.
Stars populating these orbits would contribute to the twisting of isophotes,
such as observed in the NIR, in a number of nuclear starburst galaxies
(Shaw etal. 1993; Knapen etal. 1995b). If the gas populates these orbits, 
it will quickly lose angular momentum in shocks and will fall through 
towards the nonintersecting $x_2$ orbits (if such exist), where it is expected
to accumulate. Knapen etal.  \markcite{kna95a}(1995a)
found that subsequent evolution of this gas depends crucially on its
self-gravity, which acts roughly as a surface tension. It causes the 
gaseous ring to contract by shifting to lower energy $x_2$ orbits and
ultimately shrinking across the inner ILR towards the interior 
non-intersecting $x_1$ orbits. Hence, oblique nuclear rings represent transient 
phenomena. Their characteristic time scales of $\sim10^8-10^9$ yrs and their 
potentially recurrent nature make them important when studying gas and stellar 
dynamics of galactic interiors.

Stability of accumulating gas in the ring, with its 
self-gravity increasingly dominating the local dynamics, is an important
issue.  Simulations of such gaseous rings support the above arguments and
are capable of following the gas evolution into the nonlinear regime. 
Knapen etal. \markcite{kna95a} (1995a) observed the accumulation of gas at
the inner ILR and the beginning of its subsequent collapse to the center.
This instability
is clearly dynamical, but its growth and the final outcome depends on the
efficiency of star formation and the fraction of the released stellar 
energy (winds and supernovae) which is radiated away by the ISM. Clearly, 
this phase of evolution in barred galaxies is closely related to their
circumnuclear activity and deserves further study.
An intriguing possibility is that the gas accumulated in the nuclear rings 
becomes dynamically unstable by decoupling from the background gravitational
potential, leading to a runaway instability and dumping some of this material 
much closer to the center (Shlosman, Frank and Begelman \markcite{shl89} 1989). 

In summary, we have analyzed the main periodic orbits in a barred galaxy
perturbed by a massive ring positioned in the inner resonance region, in
the vicinity of a double ILR. In some respects, the effect of the ring is
similar to that of a central spherically-symmetric mass concentration 
studied earlier in the literature, such as the secular evolution of the
stellar bar.  However, a number of profound differences exist. In
particular, we find that the orbits affected most are those exterior to the
ring, leading to the weakening and ultimate dissolution of the 
large-scale bar, while the part of the bar interior to the ring and to
the ILRs remain stable. It remains to be seen how the proposed evolution 
of bar length fits the observed difference in bar properties between 
early and late-type disk galaxies.
We also find that oblique rings are capable of
inducing an azimuthal twist in the main periodic orbits supporting the
stellar bar, similar to that observed in the NIR in some nuclear 
starburst galaxies. Overall, gas evolution in the vicinity of nuclear rings
is accelerated due an increasing number of intersecting orbits there.
This underlines the crucial role that self-gravity in the gas plays in
driving the circumnuclear activity in barred galaxies.

\acknowledgments
ACKNOWLEDGMENTS. We have benefited from discussions with 
Daniel Friedli, Marc Murison and Jerry Sellwood. This work was partially
supported by NASA grant NAGW-3839 to the University of Kentucky.

\appendix
\section{ORBIT RESPONSE TO A DOUBLE PERTURBING FORCE}

Applying epicyclic approximation to a particle on a circular
orbit with a radius $R$, we linearize its equations of motion in analogy with 
Binney \& Tremaine \markcite{bin87} (1987), i.e. the full potential
$\Phi=\Phi_0+\Phi_1$,
the orbital radius $R=R_0+R_1$, etc. Here subscript `0' refers to unperturbed
state and all perturbations are small, i.e. $\Phi_1/\Phi_0 \ll 1$ and so on.
The linearized equations of motion in polar coordinates, $R$ and $\theta$,
are given by:

\begin{equation}
d^2R_1/dt^2 + \left[\left(d^2\Phi_0/dR^2\right)-\Omega^2\right]_{R_0}
   R_1-2R_0\Omega_0d\theta_1/dt= -\left(d\Phi_1/dR\right)_{R_0},
\end{equation}
\begin{equation}
d^2\theta_1/dt^2+2\Omega_0\left(dR_1/dt\right)/R_0 = -{1\over R_0^2}
     \left[{d\Phi_1\over d\theta}\right]_{R_0}, 
\end{equation}
where $\Omega(R)\equiv\sqrt{(d\Phi_0/dR)/R}$ and $\Omega_0
\equiv \Omega(R_0)$.
Under {\it two} periodic perturbing bisymmetric forces having the same
frequency but a phase shift, e.g. of a stellar bar and of an oblique massive 
nuclear ring at an angle $\alpha$ ($\le90\arcdeg$) to the bar, the perturbing 
potential can be written as

\begin{equation}
\Phi_1(R,\theta)=\Phi_\bar(R) \cos 2\theta + \Phi_\ring(R) 
   \cos 2(\theta -\alpha), 
\end{equation}
where $\theta(t)=\theta_0(t)\equiv (\Omega_0-\Omega_\bar)t$,
in the frame of the bar and $\Phi_\ring$ is the amplitude of the
non-axisymmetric part of the ring potential. The solution to the equations
(A1-2) in the presence of the perturbing gravitational potential $\Phi_1$
can be formally written as

$$R_1(\theta_0) = {-{1\over \kappa_0^2-4(\Omega_0-\Omega_\bar)^2}}\times$$
\begin{equation}
   \left\{\left[{d\Phi_\bar\over dR}+{2\Omega\Phi_\bar\over    
   R(\Omega-\Omega_\bar)}\right]_{R_0} {\rm cos} 2\theta_0 
   + \left[{d\Phi_\ring\over dR}+{2\Omega\Phi_\ring\over    
   R(\Omega-\Omega_\bar)}\right]_{R_0} \cos 2(\theta_0-\alpha)
   \right\}.
\end{equation}

Introducing ``effective'' radial perturbing forces from a bar and a ring 
$F_{{\rm eff,}i}\equiv -d\Phi_i/dR-2\Omega\Phi_i/[R(\Omega-\Omega_\bar)]$, 
where $i=1, 2$ stand for a {\it bar} or {\it ring}, respectively, eq.(A4)
can be written as

\begin{equation}
R_1(\theta_0)={{1\over \kappa_0^2-4(\Omega_0-\Omega_\bar)^2}}
   [F_\ebar \cos 2\theta_0+F_\ering\cos 2(\theta_0-\alpha)].
\end{equation}
The orbital response to the double external force of a bar and a ring
is equivalent
to the response to a {\it single} force with an amplitude $F_{\rm eff}$, i.e.
$F_{\rm eff}^2=F_\ebar^2+F_\ering^2+2F_\ebar F_\ering \ 
\cos 2\alpha$, and with the phase $\beta$, given by

\begin{equation}
\beta = {1\over 2}\arctan{F_\ering\ \sin 2\alpha\over F_\ebar + 
    F_\ering\ \cos 2\alpha}. 
\end{equation}
The equation (A5) can therefore be written in a more compact form,

\begin{equation}
R_1(\theta_0)={{1\over \kappa_0^2-4(\Omega_0-\Omega_\bar)^2}}
  \ F_{\rm eff} \cos 2(\theta_0-\beta). 
\end{equation}
We note that $\beta$ is generally a function of $R$ and can be both positive
and negative depending on the radial change in $F_\ering(R)$ and $F_\ebar(R)$.

\end{document}